# Making the Peers' Subjective Well-being Visible Impairs Cooperator-centered Experimental Social Networks

Akihiro Nishi[1,2]*, Hiroyasu Ando[3], Meaghan Woody[4], Kamal Nayan Reddy Challa[5]

*Affiliations:*
[1] Department of Epidemiology, Fielding School of Public Health, University of California, Los Angeles (UCLA);
[2] California Center for Population Research, UCLA;
[3] Department of Biostatistics, Fielding School of Public Health, UCLA
[4] Department of Community Health Sciences, Fielding School of Public Health, UCLA;
[5] Department of Computer Science, Samueli School of Engineering, UCLA.

* Corresponding Author: Akihiro Nishi, M.D., Dr.P.H., Department of Epidemiology, Fielding School of Public Health, 650 Charles E Young Dr. S, Los Angeles, CA 90095; Tel.: +1-310-26-7140. Email: akihironishi@ucla.edu.

**Author Contributions:** AN designed the project, secured funding, and wrote the initial manuscript. AN, HA, and KNRC implemented and analyzed the experiments. AN, HA, MW, and KNRC revised the manuscript and approved the final version.
**Competing interest:** AN is a consultant to Vacan, Inc. and obtained an honorarium from Taisho Pharmaceutical Co., Ltd., which had no role in the project. KNRC is currently an employee of Alphabet Inc.
**Data, materials, and software availability:** The replication code and data will be stored in the replication package.
**Acknowledgments:** CA German provided technical support. AN is supported by the National Institutes of Health (K01AI166347), the National Science Foundation (#2230125), and the Japan Science and Technology Agency (JPMJPR21R8 and JPMJRS22K1).

**Past experiments show that reputation or the knowledge of peers' past cooperation can enhance cooperation in human social networks. On the other hand, the knowledge of peers' wealth undermines cooperativeness, and that of peers' interconnectedness and network structure does not affect it. However, it is unknown if making peers' subjective well-being (SWB) available or visible in social networks may enhance or undermine cooperation. Therefore, we implemented online network experiments (N = 662 in 50 networked groups with 15 rounds of interactions), in which study participants cooperated with or defected against connected peers through Public Goods Game, made and cut social ties with others, and rated their SWB. We manipulated the visibility of connected peers' SWB (25 visible vs. 25 invisible SWB networked groups) while keeping the connected peers' reputation and in-game wealth visible. Results show that making the peers' SWB visible did not alter overall cooperativeness, wealth, inter-connectedness, or SWB. In contrast, the visible SWB networked groups exhibited a higher number of communities and lower transitivity (the proportion of the cases where a peer of a peer is also a peer) than the invisible SWB networked groups. These phenomena are explained by an altered decision-making pattern in the visible SWB networks: cooperators were less likely to connect with cooperators and more likely to connect with defectors, and consequently, cooperators could not maintain their popularity or stay in the center of the networks.**

**Background**

Cooperation is defined as paying a cost to benefit others (Rand and Nowak, 2013). If cooperation can emerge and be sustained over social networks for the long term, it can bring prosperity to the individuals who belong to them (both peers and self); however, cooperative social networks are threatened by free-riding selfish individuals (aka defectors), who may be better off since they do not pay the cost of cooperation. Therefore, experimental economics, evolutionary biology, social physics, and other domains have investigated and reported various conditions that enhance or undermine the evolution of cooperation (Rand and Nowak, 2013) (a recent review is available (Xia et al., 2023)).

Reputation, making the information on peers' past cooperation history available or visible to self (and other participants), is reported to be a critical driver in dynamic social networks of humans (Cuesta et al., 2015; Gallo and Yan, 2015; Rand and Nowak, 2013; Xia et al., 2023), in which they may engage in reciprocal cooperation between two individuals and among a group and adjust with whom they connect. Once some individuals are labeled as defectors (i.e., with a low reputation score), they will have difficulty receiving cooperation and having social ties with cooperators. Consequently, making peers' cooperation visible leads cooperators to generate a cluster of cooperators, cut a social tie with defectors, cooperate within the cluster, and outperform defectors (Gallo and Yan, 2015). This implies that cooperators will stay in the center of cooperative social networks, while defectors may be segregated or on the periphery.

Compared to reputation, making other characteristics of peers visible has obtained less attention. In real-world human social networks, not only peers' cooperation decisions but also socioeconomic status, friendship, psychological state, and many other attributes are visible. A limited number of studies have explored this: making peers' in-game wealth visible undermines cooperation (though some studies show mixed results) (Hauser et al., 2019; Heap et al., 2016; Nishi et al., 2015); making peers' connections and network structure ("social knowledge") visible does not enhance or undermine cooperation (Corten et al., 2016; Gallo and Yan, 2015); making peers' name visible in a game setting where study participants were college classmates enhances cooperation (Wang et al., 2017). However, the role of visibility of the peers' psychological state has yet to be reported. Therefore, using an experimental platform of dynamic social networks with a cooperation game, we quantified the effect of making connected peers' SWB visible vs. invisible for cooperation, wealth, SWB, and network properties.

**Methods**

We used the breadboard.yale.edu platform and implemented a series of online network experiments (which used a shared setting as our past experiments (Nishi et al., 2023; Nishi et al., 2015)) with approval from the UCLA Office of Research Administration (#16-001920). 662 individuals were recruited online through Amazon Mechanical Turk between June 2018 and March 2019 over 50 different networked groups. They were 55.7% male, 31.4% female, and 12.8% unreported their gender, with a median age of 29 (interquartile range: 26 – 35). Any two individuals in the same networked group were initially connected by 30% at random. We randomly assigned their initial in-game wealth points, either 1,150 (30%) or 200 (70%), and thus, the expected degree of initial Gini coefficient of 0.4. Each round consisted of decision-making in PGG, rating SWB, and rewiring their social ties. In network-based PGG, study participants chose to cooperate (paying 50 points/connecting peer to benefit connected peers by providing 100 points/connecting peer) or defect (paying 0 points to benefit peers by 0 points) each round. Then, after every round, we asked study participants to rate their SWB by two questions, "How do you feel right now?" and "How would you want your neighbors to see you?" with five options (very good, good, neutral, bad, and very bad; converted from 2 to -2) (Kahneman et al., 2004; Nishi et al., 2023). In addition, study participants were given opportunities to create or cut social ties with other participants by a 30% chance (Nishi et al., 2023; Nishi et al., 2015). After cooperation and network rewiring for 15 rounds, the final in-game wealth points were paid (2,000 points = 1 USD).

While we made the connecting peers' prior-round cooperation decision-making (reputation) and in-game wealth points visible over all 50 networked groups, we manipulated the visibility of peers' SWB. Only in the visible SWB condition (25 networked groups), smiley-to-non-smiley face emojis were shown on the study participants' web screen, which represent their answer to the second SWB question **(Fig.1)**. The use of the emoji was derived from the emoji-based visual analog scale in the health care setting (He et al., 2022).

For data analysis, we used R *lme4* package (ver.1.1-35.3) and *lmerTest* package (ver.3.1-3) to construct regression models that took into account the hierarchical data structure, in which observations were clustered over study participants, networks, and rounds. We also used R *igraph* package (ver.2.0.3) for social network analysis and *mediation* package (ver.4.5.0) for mediation analysis.

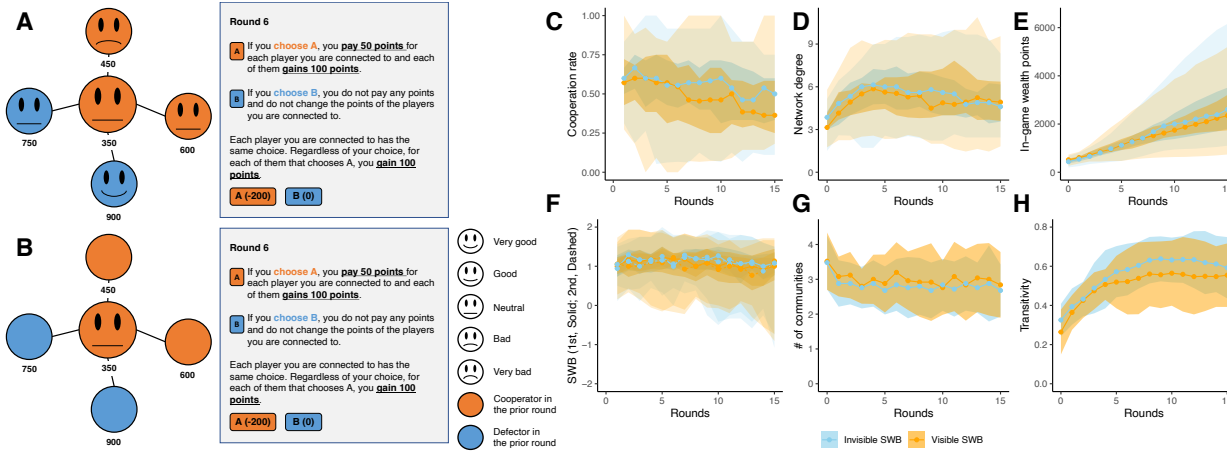

**Figure 1. Experimental social networks in our set-up for the visible subjective well-being (SWB) condition (A) and the invisible SWB condition (B) and the trajectories of six outcome variables in the visible and invisible SWB conditions (C-H). C-F.** The lines with dots represent the medians. The shades represent interquartile ranges, minimum, and maximum. **G-H.** The lines with does represent means. The shades represent ±1 s.e.m.

## Results

*Effects of the visible SWB condition.* Making the peers' SWB visible vs. invisible was not associated with the overall cooperation rate (49.3% vs. 53.3%, $P = 0.570$), network degree (the number of connected peers: 5.47 vs. 5.58, $P = 0.808$), in-game wealth points (1,558 vs. 1,588, $P = 0.715$), or SWB (1.02 vs. 1.10, $P = 0.517$ for the first question; 0.96 vs. 1.07, $P = 0.314$ for the second question) **(Figs.1C-F)**. On the other hand, the visible vs. invisible SWB condition exhibited a higher number of communities (2.77 vs. 2.58, $P = 0.051$), which was detected by the Louvain method in *igraph*, and lower transitivity (the proportion of two connected peers being also connected: 0.505 vs. 0.549, $P = 0.054$) **(Figs.1G-H)**. The results imply that how study participants rewire their social ties with others may systematically differ between the visible and invisible SWB conditions.

*Decision-making mechanisms to explain the effects.* The investigation that focused on the difference in the rewiring decision-making at the individual level identified that, in the *invisible* SWB condition, cooperators in the current round chose to connect with cooperators by 86.1% and defectors by 29.5%; in the *visible* SWB condition, cooperators chose to connect with cooperators by 82.0% (less frequent homophilic connections by cooperators) and defectors by 30.3% (more frequent heterophilic connections by cooperators). In addition, in the *invisible* SWB condition, defectors in the current round chose to connect with defectors by 60.5% and cooperators by 72.1%; in the *visible* SWB condition, defectors chose to connect with defectors by 56.4% (less frequent homophilic connections by defectors) and cooperators by 78.2% (more frequent heterophilic connections by defectors). In sum, regardless of the cooperation decision-making of focal individuals, making peers' SWB visible reduces homophilic tie formation and amplifies heterophilic tie formation (interaction $P = 2.4 \times 10^{-9}$). Consequently, cooperators' triangles (a triad made by three cooperators) were less likely to exist in the visible SWB condition (6.45% vs. 8.53% among all possible triangles of any three participants randomly selected from the same networked group), which can explain the reduction in transitivity in the visible SWB condition.

The eigenvector centrality of cooperators was much higher than that of defectors in the *invisible* SWB condition (0.760 vs. 0.609, $P < 2.0 \times 10^{-16}$), which suggests that cooperators stayed in the center and defectors stayed on the periphery of the networks. A high eigenvector centrality means that a focal individual connects with many individuals of high eigenvector centrality. On the other hand, in the *visible* SWB condition, the discrepancy was attenuated: the eigenvector centrality of cooperators decreased, and that of defectors increased (0.711 vs. 0.633, $P = 2.9 \times 10^{-10}$; interaction $P = 8.4 \times 10^{-3}$). This implies that cooperators are less likely to stay in the center of social networks in the visible SWB condition.

***Mediation analysis.*** The effect of the visible/invisible SWB condition (x) for the number of communities (y) was mediated by the average eigenvector centrality of cooperators (z) through a mediation path (x→z→y, 68.8%, $P = 0.004$). That for transitivity (as y) was also mediated by the average eigenvector centrality of cooperators (42.3%, $P < 2.0 \times 10^{-16}$). The results suggest that the network locations of cooperators (and defectors) play a critical role in determining the number of communities and transitivity of each networked group.

**Discussion**

Our experiments revealed that cooperators' embedding over dynamic social networks differed when connected peers' SWB was made visible versus invisible. In the invisible SWB condition, cooperators could keep connected with other cooperators, maintain their popularity, and stay close to the center of the networked groups. In contrast, in the visible SWB condition, cooperators were dispersed over the networks, and consequently, the networked groups were more segregated (i.e., a higher number of the detected communities). Suppose we aim to construct cooperator-centered social networks, for example, in workplaces or schools. Our experiments suggest that making peers' happiness and unhappiness fully visible or disseminated can be problematic.

In addition, we have confirmed that emojis representing SWB cannot be an alternative to reputation, which enhances cooperation (Cuesta et al., 2015; Gallo and Yan, 2015; Rand and Nowak, 2013; Xia et al., 2023). Making peers' SWB visible did not alter the cooperation, network degree, wealth, or SWB. Although the visible SWB condition exhibited a slightly lower cooperation rate (49.3% vs. 53.3%) **(Fig.1C)**, it is not at the detectable level.

In the real world, peers' SWB is somewhat visible. Although proving this claim and measuring the visibility of peers' SWB is challenging, it is known that people's happiness and depression spread over social networks (Christakis and Fowler, 2013), which can happen only when peers' SWB is visible. One future research direction could be to find the optimal level of the visibility of peer's SWB. For example, mass media do not sometimes broadcast celebrities' suicides to prevent copycat suicidal attempts (Cheng et al., 2007); YouTube has made the number of "dislikes" that each video received invisible since 2021 (The YouTube Team, 2021). During the COVID-19 pandemic, using a face mask made reading peers' facial expressions difficult, which brought severe concern in child development and other domains (Mheidly et al., 2020). Our experiment aimed to mimic these real-world examples of manipulating the visibility of peers' SWB and finding its optimal level in the experimental setting. However, a crucial limitation is that laboratory experiment results, including ours, cannot directly or immediately translate into real-world policymaking.